\documentstyle[12pt]{article}

\def\lsim{\mathrel{\lower2.5pt\vbox{\lineskip=0pt\baselineskip=0pt
           \hbox{$<$}\hbox{$\sim$}}}}
\def\gsim{\mathrel{\lower2.5pt\vbox{\lineskip=0pt\baselineskip=0pt
           \hbox{$>$}\hbox{$\sim$}}}}

\begin{document}

\setlength{\baselineskip}{8mm}

\begin{titlepage}

\begin{flushright}
DPNU-94-37 \\
AUE-06-94 \\
hep-ph/9409361 \\
Revised Version \\
\end{flushright}

\vspace{5mm}

\begin{center}
{\large  \bf
The Aligned $SU(5) \times U(1)^2$ Model
}

\vspace{15mm}

Naoyuki HABA$^1$, Chuichiro HATTORI$^2$,
Masahisa MATSUDA$^3$, \\
Takeo MATSUOKA$^1$  and Daizo MOCHINAGA$^1$ \\
{\it
${}^1$Department of Physics, Nagoya University \\
           Nagoya, JAPAN 464-01 \\
${}^2$Science Division, General Education \\
     Aichi Institute of Technology \\
      Toyota, Aichi, JAPAN 470-03 \\
${}^3$Department of Physics and Astronomy \\
     Aichi University of Education \\
      Kariya, Aichi, JAPAN 448 \\
}

\end{center}

\vspace{10mm}

\begin{abstract}

In Calabi-Yau string compactification,
it is pointed out that there exists
a new type of $SU(5) \times U(1)^2$ model
(the aligned $SU(5) \times U(1)^2$ model)
in which the $SU(5)$ differs from
the standard $SU(5)$
and also from the flipped $SU(5)$.
With the aid of the discrete symmetry suggested from
Gepner model, we construct a simple and
phenomenologically interesting three-generation model
with the aligned $SU(5) \times U(1)^2$ gauge symmetry.
The triplet-doublet splitting problem can be solved.
It is also found that there is a realistic solution
for solar neutrino problem and for the $\mu $-problem.
At low energies this model is in accord with the minimal
supersymmetric standard model except for the existence
of singlet fields with masses of $O(1)$TeV.

\end{abstract}

\end{titlepage}

\section{Introduction}

It is very plausible that the Planck scale$(M_{\rm Pl})$
is the fundamental scale of the theory
which unifies all fundamental interactions.
The only known candidate of the consistent Planck
scale theory is the heterotic superstring theory.
On the other hand, the standard model is consistent
with many of observations at low energies.
How does the superstring theory connect with
the standard model ?
How does the hierarchical ramification of
the unified interaction occur ?
Especially, it is important to clarify the energy scale
of the ramification into $SU(3)_c$  and
$SU(2)_L$ gauge interactions.
If $SU(3)_c$  and $SU(2)_L$ gauge interactions are unified
at the Planck scale,
the ramification must have its origin in the flux breaking
associated with the multiply-connectedness of
the compactified manifold.
If we have GUT types of gauge group such as
$SU(5), SO(10)$ at the scale smaller than $M_{\rm Pl}$,
the ramification into $SU(3)_c$ and $SU(2)_L$ needs to occur
at an intermediate energy scale through Higgs mechanism.
The scale of the ramification into $SU(3)_c$  and
$SU(2)_L$ is closely related to the longevity of proton.
For superstring models to be consistent with
proton stability,
it is required that $SU(3)_c$-triplet
and $SU(2)_L$-doublet gauge bosons
in ${\bf 78}$-representation of $E_6$ get
masses of $O( \gsim 10^{16})$GeV.
On the other hand, it is commonly
considered that in superstring models
Higgs mechanism can hardly occur at a scale
of $O( \gsim 10^{16})$GeV.
For this reason, until now many authors have
preferred the case in which
the ramification into $SU(3)_c$ and $SU(2)_L$
is due to flux breaking at the Planck scale.
However, if there appear mirror chiral superfields in
the effective theory and if an appropriate discrete
symmetry restricts nonrenormalizable interactions
to a special form,
it is theoretically possible that Higgs mechanism
occurs at a scale
of $O( \gsim 10^{16})$GeV
\cite{discrete}.

The purpose of this paper is to study the GUT type
scenario with $SU(5)$ gauge symmetry
in Calabi-Yau string compactification.
In this scenario Higgs mechanism should occur at a scale
$M_X$ with $M_{\rm Pl} > M_X \gsim 10^{16}{\rm GeV}$.
As a result, we find a new type of $SU(5) \times U(1)^2$
model, which is named the aligned $SU(5) \times
U(1)^2$ model by the reason shown later.
As is well known, there is a disparity
between $M_{\rm Pl}$ and the unification scale $O(10^{16})$GeV
of gauge couplings in the minimal supersymmetric standard
model
\cite{Amaldi}.
In the scenario with the aligned $SU(5) \times U(1)^2$
it is possible to solve such a disparity.
In this paper we construct a realistic three-generation
model with the aligned $SU(5) \times U(1)^2$ gauge symmetry.
In the model $SU(3)_c$ and $SU(2)_L$ gauge couplings
come together at the scale $O(10^{17.5})$GeV,
while the aligned $SU(5)$ and $U(1)^2$ gauge interactions
are unified at the Planck scale.

In the four-dimensional effective theory from
Calabi-Yau compactification the gauge
symmetry $G$ at the Planck scale
becomes a subgroup of $E_6$.
Phenomenologically it is required that
the standard gauge group
$G_{st}=SU(3)_c \times
SU(2)_L \times U(1)_Y$ is contained in $G$.
As an example of GUT types of the $G$ there is
an $SU(5) \times U(1)^2$ group.
When we embed $G_{st}$ into $SU(5) \times U(1)^2$,
we obtain different types of $SU(5)$
according as the $SU(5)$ entirely contains
$U(1)_Y$ or not.
On the other hand, we assign matter fields to {\bf 27}
of $E_6$ so as to connect the effective theory with
the standard model.
In the standard model $Y$-charges are settled for
$SU(2)_L$-doublet superfields $Q$, $L$ of quarks
and leptons, singlet superfields $U^c$, $D^c$, $E^c$
and Higgs-doublet superfields $H_u$, $H_d$.
Then $U(1)_Y$ should be embedded into $E_6$
so that we can reproduce $Y$-charges of these
matter fields.
Furthermore, it is plausible for us to require that
in the effective theory there appear the Yukawa
interactions $QU^cH_u$, $QD^cH_d$, $LN^cH_u$ and $LE^cH_d$
to get Dirac masses of quarks and leptons,
where $N^c$ represents a superfield of conjugate neutrino.
We study GUT types of model
under these constraints on the effective theory.
For illustration we take up an $SU(5) \times U(1)^2$
gauge group.

In the case $U(1)_Y \subset SU(5)$
quark and lepton superfields in
${\bf 27}$-representation of the $E_6$ belong to ${\bf 10}$
and ${\bf 5^*}$ representations of $SU(5)$ as
\[
\begin{array}{clll}
   {\bf 10}:  & Q,   & U^c, & E^c,  \\
   {\bf 5^*}: & D^c, & L.   &
\end{array}
\]
The $SU(5)$ of this case is the standard $SU(5)$
\cite{SU5G}.
Hereafter we denote this $SU(5)$ as $SU(5)_S$.
In superstring models, however, we have no Higgs
superfields in an adjoint representation.
Therefore, the $SU(5)_S$ symmetry can not
be broken spontaneously into the standard gauge group
$G_{st}$ through Higgs mechanism
\cite{gaugesym1}
\cite{gaugesym2}.
Thus GUT type of models with $SU(5)_S$ are
excluded in the scheme of Calabi-Yau compactification.

In the case $U(1)_Y \not \subset SU(5)$
we have two different types of the assignment
of matter fields.
The situation is as follows.
In addition to the above-mentioned matter fields,
in {\bf 27} of $E_6$ we have an extra $G_{st}$-neutral
superfields $S$ which is a standard $SO(10)$-singlet,
and extra colored superfields $g, g^c$.
Among these matter fields
$(D^c, g^c)$, $(L, H_d)$ and $(S, N^c)$
are indistinguishable with respect to $G_{st}$,
respectively.
Therefore, at first sight it seems that
we may interchange the assignment of these fields
to the {\bf 27} states at will.
However, under the requirement that we get the Yukawa
interactions $LN^cH_u$ and $QD^cH_d$,
it is only possible for us to interchange these fields
in sets of $(D^c, L, S)$ and $(g^c, H_d, N^c)$.
These assignments implies that extra colored superfields
$g$ and $g^c$ mediate proton decay and then
hereafter we refer $g$ and $g^c$ as leptoquark superfields.
These leptoquark fields get masses through the Yukawa
interaction $Sg^cg$ with a non-zero VEV of $S$.
Depending on whether the $S$ resides in ${\bf 10}$ or
${\bf 1}$ of $SU(5)$,
we have two different types of $SU(5)$ in the case
$U(1)_Y \not \subset SU(5)$.
In the case that the $S$ belongs to {\bf 10} of $SU(5)$,
a non-zero VEV of $S$ results in the spontaneous
breaking of the $SU(5)$ symmetry.
While, in the case the $S$ resides in {\bf 1} of $SU(5)$,
the $SU(5)$ is unbroken even with a non-zero VEV of $S$.

In the case that the $S$ belongs to {\bf 1} of $SU(5)$,
quarks and leptons are assigned
as
\[
\begin{array}{clll}
   {\bf 10}:  & Q,   & D^c, & N^c,  \\
   {\bf 5^*}: & U^c, & L,   &       \\
   {\bf 1}:   & E^c. &      &
\end{array}
\]
This assignment of matter fields to the representations
of $SU(5)$ is the same in the case of the
flipped $SU(5) \times U(1)$ model
\cite{flipped1}\cite{flipped2}.
Then we denote the $SU(5)$ of this case as $SU(5)_F$.
The so-called flipped $SU(5) \times U(1)$ model
is derived from the compactification
in which the holonomy group is $SO(6)$
\cite{flipped3}.
On the other hand, in Calabi-Yau compactification
there is a possibility of
the flipped $SU(5) \times U(1)^2$ model.
An extra $U(1)$ $(U(1)_\psi )$ gauge symmetry
distinguishes the flipped $SU(5) \times U(1)^2$ model
from the flipped $SU(5) \times U(1)$ model.
{}From the study of mass spectra it turns out that
the flipped $SU(5) \times U(1)^2$ model is
not realistic.

The case of $SU(5)$'s that the $S$ resides in {\bf 10}
of $SU(5)$ is a new type of $SU(5)$.
In this case matter fields are assigned as
\[
\begin{array}{clll}
   {\bf 10}:  & Q,   & g^c, & S,   \\
   {\bf 5}:   & L,   & g,   &      \\
   {\bf 5^*}: & D^c, & H_u, &      \\
   {\bf 5^*}: & U^c, & H_d, &      \\
   {\bf 1}:   & E^c, &      &      \\
   {\bf 1}:   & N^c. &      &
\end{array}
\]
In this case quark and lepton superfields belong
separately to six irreducible representations
and {\it are aligned in the front row on the above list}.
Then a new type of $SU(5)$ is named the aligned $SU(5)$
and denoted as $SU(5)_A$.
This type of GUT model has been first discussed by
Panagiotakopoulos
\cite{Pana},
who studied $SU(6) \times U(1)$ models constructed using
the Tian-Yau manifold divided by $Z_3$.
However, in Ref\cite{Pana} down-type quarks, lepton-doublet
and right-handed neutrinos are denoted as $g^c$, $H_d$ and
$S$, respectively.
This is due to the flipped type of assignment of matter fields,
in which assignment the $SU(6)$ contains the flipped $SU(5)$.

This paper is organized as follows.
In section 2 we briefly review the relation between
flux breaking and gauge symmetry at the Planck scale
and then carry out the classification of the gauge
groups.
It is shown that through the abelian flux breaking
there possibly appear three kinds of $SU(5) \times
U(1)^2$ gauge symmetry as mentioned above.
Among them the aligned $SU(5) \times U(1)^2$ model
can be consistent with proton stability,
when $\langle S \rangle \gsim 10^{16}$GeV.
In section 3 we find gauge hierarchies
for four types of models
and clarify the processes
of symmetry breaking to $G_{st}$.
By introducing an appropriate discrete symmetry suggested
from Gepner model in section 4,
we construct a simple three-generation model
with the aligned $SU(5) \times U(1)^2$ gauge symmetry
and discuss its phenomenological implication.
In the model the generation and the anti-generation
numbers are 4 and 1, respectively.
It is pointed out that there is an interesting
solution for the triplet-doublet splitting
problem and for solar neutrino problem.
At low energies this model is
in line with the minimal supersymmetric standard
model except for the existence of $G_{st}$-singlet
superfields.
Section 5 is devoted to summary and discussion.
We also show that there is a realistic solution
for the $\mu $-problem.


\section{Flux breaking mechanism}

In Calabi-Yau compactification on
multiply-connected manifold $K$
there generally exists a nontrivial
Wilson loop $U$ on $K$ and then the available gauge group
$G$ at the Planck scale is reduced to a subgroup of $E_6$.
The nontrivial $U$ gives rise to the discrete
symmetry ${\overline G_d}$,
which is an embedding of $G_d = \Pi _1(K)$ in the $E_6$.
Then $G$ consists of the generators of $E_6$ which commute
with all elements of ${\overline G_d}$.
This mechanism is called flux breaking or Hosotani
mechanism
\cite{Hosotani}.
Phenomenologically it is required that
the group $G$ contains $G_{st}$.
The generators of $E_6$ are denoted as \{ $H_i$, $E_{\xi}$ \}
in Cartan-Weyl basis,
where $H_i$'s are diagonal generators and $E_{\xi}$'s
are ladder operators associated with root vectors $\xi $.
In the abelian flux breaking the Wilson loop $U$
is expressed as
%
\begin{equation}
     U = \exp(2\pi i\sum_i z_iH_i),
\end{equation}
%
where $z_i$'s are real parameters.
In this case we obtain
%
\begin{equation}
     U E_{\xi} U^{-1}= \exp \left\{ 2\pi i (Z, \xi )
                                      \right\} \,E_{\xi},
\end{equation}
%
where
%
\begin{equation}
        (Z, \xi) = \sum_i z_i \xi_i , \ \ \ \ \
        [H_i, E_{\xi}] = \xi _i E_{\xi }.
\end{equation}
%
Under the condition $G_{st} \subset G$
the vector $Z$ is described in
terms of three real parameters $\alpha ,\beta, \gamma$
as
\cite{gaugesym2}
%
\begin{equation}
   Z = \alpha \ \Theta_1 + \beta \ \Theta_2 + \gamma \ \Theta_3,
\end{equation}
%
where $\Theta_i (i=1,2,3)$ stand for three linearly
independent $SU(3)_c \times SU(2)_L$-neutral weights
in ${\bf 27}$ representation and
coincide with weights of
$E^c$, $S$, $N^c$, respectively.
The relations $(\Theta_i, \Theta_j)
= \delta_{ij} + 1/3 $ hold.
When $U(1)_Y$ is embedded into $G$,
$U(1)_Y$-generator is
%
\begin{equation}
      Y = \frac {1}{3} ( 5 \Theta_1 -\Theta_2 -\Theta_3 ),
\end{equation}
%
which is orthogonal to $\Theta_{2,3}$.

Among gauge bosons in the ${\bf 78}$ of $E_6$,
there are three sets of $({\bf 3, 2})$ gauge bosons
with respect to $SU(3)_c \times SU(2)_L$
\cite{gaugesym2}.
We denote three representatives of root vectors
corresponding to these three sets of gauge bosons
as $\xi^{\rm (A)}, \xi^{\rm (B)}, \xi^{\rm (C)}$.
The quantum numbers of these root vectors
are
%
\begin{eqnarray}
   \xi^{\rm (A)} \ \ &:& \ \
                  ({\bf 3, 2}, -5/3, 0,  0), \nonumber \\
   \xi^{\rm (B)} \ \ &:& \ \
                  ({\bf 3, 2},  1/3, -1,  -1), \\
   \xi^{\rm (C)} \ \ &:& \ \
                  ({\bf 3, 2},  1/3, 1, -1)   \nonumber
\end{eqnarray}
%
under $SU(3)_c \times SU(2)_L \times U(1)_Y \times
U(1)_I \times U(1)_{\eta }$ symmetry,
where $U(1)_I$ and $U(1)_{\eta }$ correspond to
$\Theta_2 \mp \Theta_3$ axes, respectively.
Referring to $U(1)_Y$-charges we can discriminate
$\xi^{(A)}$ from $\xi^{(B)}$ and $\xi^{(C)}$.
On the other hand, $\xi^{(B)}$ and $\xi^{(C)}$
are indistinguishable with respect to $G_{st}$.
Inner products of these root vectors with $Z$
become
%
\begin{equation}
  (Z, \xi^{\rm (A)}) = - \alpha,
              \ \ (Z, \xi^{\rm (B)}) = - \beta,
                    \ \ (Z, \xi^{\rm (C)}) = - \gamma.
\end{equation}
%
The gauge group $G$ at the Planck scale is
determined depending on values of these parameters
$\alpha, \beta, \gamma $.
{}From Eq.(2), if we obtain
%
\begin{equation}
(Z,\xi) \equiv 0  \ \ \ \ \ {\rm mod} \ 1,
\end{equation}
%
then the $E_{\xi}$ becomes a generator of $G$.
To the contrary, if we get
%
\begin{equation}
(Z,\xi) \not\equiv 0  \ \ \ \ \ {\rm mod} \ 1,
\end{equation}
%
the $E_{\xi}$ is not a generator of $G$.
In the case $\alpha, \beta, \gamma \not\equiv 0$
we do not have any kinds of $SU(5)$ symmetry
and all three kinds of $({\bf 3, 2})$ gauge boson
become massive at $O(M_{\rm Pl})$
\cite{gaugesym2}.
On the other hand, when one of $\alpha, \beta, \gamma $
becomes zero (mod 1),
there appear three kinds of $SU(5)$ symmetry
as
%
\begin{eqnarray}
    {\rm (A)} \ \ \alpha \equiv 0, \ \ \beta,
                     \ \gamma \not\equiv 0 \ \ \
                  &:& \ \ \ G \supset SU(5)_S \times U(1)^2, \\
    {\rm (B)} \ \ \beta  \equiv 0, \ \ \gamma,
                     \ \alpha \not\equiv 0 \ \ \
                  &:& \ \ \ G \supset SU(5)_A \times U(1)^2, \\
    {\rm (C}) \ \ \gamma \equiv 0, \ \ \alpha,
                     \ \beta \not\equiv 0 \ \ \
                  &:& \ \ \ G \supset SU(5)_F \times U(1)^2.
\end{eqnarray}
%
Here $U(1)^2$-axes in these cases correspond to
$\Theta_2 \mp \Theta_3$, $\Theta_3 \mp \Theta_1$
and $\Theta_1 \mp \Theta_2$, respectively.
As will be discussed in the next section,
depending on the relations between two nonzero
parameters the gauge group $G$ varies from
$SU(5) \times U(1)^2$ to $SU(6) \times SU(2)$
and is classified into four cases.
For a moment, we concentrate on the case
$G = SU(5) \times U(1)^2$.
In a {\bf 27} representation of $E_6$,
there are two sets of $(D_1,L_1,S_1)$ and $(D_2,L_2,S_2)$.
$D_{1,2}$ reside in the representation of
$({\bf 3}^*,{\bf 1},{2\over3})$
as to $(SU(3)_c,SU(2)_L,U(1)_Y)$.
Similarly, $L_{1,2}$ and $S_{1,2}$ reside
in $({\bf 1},{\bf 2},-{1\over2})$
and $({\bf 1},{\bf 1},0)$, respectively.
$(D_1,L_1,S_1)$ have positive $U(1)_I$-charges
and $(D_2,L_2,S_2)$ have negative $U(1)_I$-charges, respectively.
A phenomenologically viable model implies that
three generations of down-type quarks
$({\bf 3}^*,{\bf 1},{2\over3})$ (denoted as $D^c$)
remain massless at TeV scale.
Thus the other $({\bf 3}^*,{\bf 1},{2\over3})$
field in {\bf 27} is considered as the leptoquark
(denoted as $g^c$).
We can assign $D_1$ to down-type quarks $D^c$
without the loss of generality.
Then the sign of $U(1)_I$-charges is fixed.
One can write down eleven $E_6$-invariant Yukawa couplings
by using ${\bf 27}$ fields as
\[
\begin{array} {llll}
g S_1 D_2, &  g S_2 D_1, &  QD_1L_2,   & QD_2L_1,   \\
H_uL_1S_2, &  H_uL_2S_1, &  E^cL_1L_2, & U^cD_1D_2, \\
QQg,       &  QU^cH_u,   &  U^cE^cg.   &
\end{array}
\]
Now it is phenomenologically plausible for us
to require the existence of
the Yukawa couplings $QD_1L_2(=QD^cH_d)$
and $E^cL_1L_2(=E^cLH_d)$
to obtain available Dirac masses of quarks and
leptons at weak scale.
Since we take $D_1$ as down-quark $D^c$,
we should assign $L_2$ to $H_d$ and $L_1$ to $L$, respectively.
Moreover, the  coupling $H_uL_1S_2$ can give the neutrino
Dirac masses through $\langle H_u\rangle$
so that we should assign $S_2$ to right-handed neutrino $N^c$.
Consequently, under the assignment of $D_1=D^c$
we must take $(D_1,L_1,S_1)$ as $(D^c,L,S)$,
which have positive $U(1)_I$-charges,
and also $(D_2,L_2,S_2)$ as $(g^c,H_d,N^c)$,
which have negative $U(1)_I$-charges.
We denote here $S_1$ as $S$.
For three cases (A), (B) and (C) irreducible decompositions
of the ${\bf 27}$ matter fields under $SU(5) \times U(1)^2$
are shown in Table I.
In the case (A) both $S$ and $N^c$ reside in ${\bf 1}$
of $SU(5)$.
While, in the case (B) $S$ resides in ${\bf 10}$ of $SU(5)$
but $N^c$ in ${\bf 1}$.
In the case (C) $N^c$ resides in ${\bf 10}$ of $SU(5)$
but $S$ in ${\bf 1}$.

%
\vskip 0.5cm

\begin{center}
 \unitlength=0.8cm
 \begin{picture}(2.5,2.5)
  \thicklines
  \put(0,0){\framebox(3,1){\bf Table I}}
 \end{picture}
\end{center}

\vskip 0.5cm
%

The $SU(5)_S$ in the case (A) is just the standard $SU(5)$
\cite{SU5G}.
In order to break down $SU(5)_S$ into $G_{st}$
at an intermediate energy scale,
$({\bf 3, 2}, -5/3)$ gauge superfields under
$SU(3)_c \times SU(2)_L \times U(1)_Y $ should become
massive via Higgs mechanism.
However, there are no $({\bf 3, 2}, -5/3)$ chiral superfields
in the {\bf 27} which would be absorbed to give masses
to $({\bf 3, 2}, -5/3)$ gauge superfields.
Thus in the case (A) we can not construct a realistic model.
The standard $SU(5)$-GUT model is excluded
in the Calabi-Yau string theory.

The $SU(5)_A$ in the case (B) can be broken into
$SU(3)_c \times SU(2)_L$,
when $S$ develops a nonzero VEV.
This is due to the fact that $S$ belongs to
the ${\bf 10}$ of $SU(5)_A$.
When we decompose $SU(5)_A$ into
$SU(3)_c \times SU(2)_L \times U(1)_{Y'}$,
we obtain
%
\begin{equation}
        Y' = \frac {1}{3} (5\Theta_2 - \Theta_3 - \Theta_1).
\end{equation}
%
The $U(1)_Y$ is given by a linear combination of
$U(1)_{Y'}$ and $U(1)^2$ aside from $SU(5)_A$.
In this case a quark-doublet superfield $Q$ is absorbed by
$({\bf 3, 2}, 1/3)$ gauge superfields via Higgs mechanism.
The $({\bf 3, 2}, 1/3)$ gauge superfields gain masses
of order $\langle S \rangle $.
Proton decay is caused not only by the interactions
of $({\bf 3, 2}, 1/3)$ gauge superfields
but also by the interactions of leptoquark superfields
$g$ and $g^c$.
In the $SU(5)_A \times U(1)^2$ model
we have four independent Yukawa coupling constants
$\lambda ^{(r)}\ (r = 1 \sim 4)$ which appear
in the superpotential
%
\begin{eqnarray}
      W_Y & = & \lambda ^{(1)} \left(
                  QQg + Qg^cL + g^cSg
                            \right) \nonumber \\
          &   & \quad + \lambda ^{(2)} \left(
                   QU^cH_u + QH_dD^c + g^cU^cD^c + SH_dH_u
                            \right) \nonumber \\
          &   & \quad + \lambda ^{(3)} \left(
                   LH_dE^c + gU^cE^c
                            \right)
                      + \lambda ^{(4)} \left(
                   LH_uN^c + gD^cN^c
                            \right),
\label{eqn:WY}
\end{eqnarray}
%
where the generation indices are omitted
and $\lambda $'s are all expected to be $O(1)$.
{}From Eq.(\ref{eqn:WY}) leptoquark superfields $g$
and $g^c$ also gain masses of the order
$\langle S \rangle $ through the Yukawa interactions
$\lambda ^{(1)}g^cSg$.
Thus, at energies below $\langle S \rangle $,
$g$ and $g^c$ decouple from the effective theory.
Therefore, if $\langle S \rangle $ is equal to or
larger than $O(10^{16})$GeV,
this model is consistent with proton
stability.

The flipped type of $SU(5)$
in the case (C) can be decomposed into
$SU(3)_c \times SU(2)_L \times U(1)_{Y''}$.
The generator $Y''$ is given by
%
\begin{equation}
        Y'' = \frac {1}{3} (5\Theta_3 - \Theta_1 - \Theta_2).
\end{equation}
%
When $N^c$ develops a nonzero VEV,
a quark-doublet superfield $Q$ is absorbed
to give masses of $O(\langle N^c \rangle )$
to $({\bf 3, 2},1/3)$ gauge superfields.
In the $SU(5)_F \times U(1)^2$ model Yukawa interactions
are given by
%
\begin{eqnarray}
      W_Y & = & \lambda ^{(1)} \left(
                  QQg + QD^cH_d + D^cN^cg
                            \right) \nonumber \\
           &   & \quad + \lambda ^{(2)} \left(
                   QH_uU^c + Qg^cL + D^cg^cU^c + N^cH_uL
                            \right) \nonumber \\
           &   & \quad + \lambda ^{(3)} \left(
                   H_uH_dS + g^cgS
                            \right)
                + \lambda ^{(4)} \left(
                   LH_dE^c + U^cgE^c
                            \right).
\end{eqnarray}
%
In the symmetry breaking due to a nonzero
$\langle N^c \rangle $,
leptoquark superfields $g$ and $g^c$ can not gain
masses of $O(\langle N^c \rangle )$.
Unless $S$ develops a large VEV,
we are led to the fast proton decay.
To avoid this difficulty,
$S$ also should develop its VEV of $O(\gsim 10^{16})$GeV.
Thus in the case (C)
it is required that both $\langle S\rangle $ and
$\langle N^c\rangle $ are $O(\gsim 10^{16})$GeV.
In this scheme of symmetry breaking it is impossible
for us to get a large Majorana-mass of right-handed neutrino
\cite{majom}.
Furthermore, since the Yukawa couplings of $g^cgS$ and
$H_uH_dS$ take a common value $\lambda ^{(3)}$,
we can not solve the triplet-doublet splitting problem.
Thus the $SU(5)_F \times U(1)^2$ model
is not realistic.
It should be noted that the present $SU(5)_F \times
U(1)^2$ model is quite different from the
flipped $SU(5) \times U(1)$ model.
The $U(1)$ factor group in the flipped $SU(5) \times
U(1)$ model corresponds to $(4\Theta _1 - \Theta _2)/2$,
while an extra $U(1)$ symmetry in the $SU(5)_F \times
U(1)^2$ model relative to the flipped $SU(5) \times
U(1)$ model corresponds to $\Theta _2$-axis,
i.e. $U(1)_{\psi }$.
Due to the extra $U(1)_{\psi }$ symmetry,
many of the Yukawa interactions such as
$({\bf 10^*\cdot 10^*\cdot 5^*})$,
$({\bf 10\cdot 10^*\cdot 1})$
which appear in the flipped $SU(5) \times
U(1)$ model are forbidden.
As a consequence, unlike in the
flipped $SU(5) \times U(1)$ model
the triplet-doublet splitting mechanism and
see-saw mechanism
\cite{seesaw}
\cite{flipped4}
 are not
at work in the $SU(5)_F \times U(1)^2$ model.
Furthermore, there is a sharp distinction
between the present $SU(5)_F \times U(1)^2$ model
and the flipped $SU(5)_F \times U(1)$ model
with respect to their generation structure.

When all but one of $\alpha, \beta, \gamma $ are zero (mod 1),
there appears an $SO(10) \times U(1)$ gauge group.
In this case we also have three kinds of model.
For instance, in the case
$\alpha, \gamma \equiv 0$ and $\beta \not\equiv 0$
the $SO(10)$ referred here is the same as the usual one.
As mentioned above, not only the case $\alpha \equiv 0$
but also the case $\gamma \equiv 0$
are unfavorable.
Thus we have no possibilities of the $SO(10) \times U(1)$
gauge symmetry.

Next we consider the case of non-abelian flux breaking.
Root vectors of the $E_6$ perpendicular to those of $G_{st}$
are restricted only to $\pm(\Theta_2 - \Theta_3)$.
Since these root vectors compose $SU(2)$ group,
the remaining gauge symmetry is at most $SU(6)$.
This $SU(6)$ involves $SU(5)_S$ but neither $SU(5)_A$
nor $SU(5)_F$.
Since we have no realistic solutions for the $SU(5)_S$-GUT,
$SU(3)_c$ and $SU(2)_L$ should be already separated
in the non-abelian flux breaking at the Planck scale.


\section{ Gauge hierarchies }

As discussed in the previous section,
the realistic scenarios with $SU(5) \times U(1)^2$
gauge symmetry are limited only
to the case (B) of flux breaking
%
\begin{equation}
    \beta \equiv 0, \ \ \ \gamma, \ \alpha \not\equiv 0.
\end{equation}
%
Then we proceed to study the case (B)
including the aligned $SU(5) \times U(1)^2$ model.
To explain large Majorana-masses
of right-handed neutrinos,
we now consider the hierarchical
symmetry breaking with $\langle S \rangle
\gsim 10^{16}{\rm GeV} \gg \langle N^c \rangle \gg
m_{\rm susy}$
\cite{Masip}
\cite{majom},
where $m_{\rm susy}$ represents the susy breaking scale
of $O(1)${\rm TeV}.
Some of Gepner models
\cite{Gepner}
 potentially implement
this hierarchical type of symmetry breaking.

To maintain supersymmetry down to $m_{\rm susy}$,
the $D$-terms should vanish at
large scales $\langle S \rangle $ and
$\langle N^c \rangle $.
It is realized by assuming the existence of mirror
chiral superfields of $S$ and $N^c$ and by setting
$\langle S \rangle = \langle {\overline S} \rangle $ and
$\langle N^c \rangle = \langle {\overline N^c} \rangle $.
In what follows we take up this scheme of
symmetry breaking.
The gauge group $G$ in the region ranging from
the Planck scale to the scale $\langle S \rangle $
is classified into four cases
depending on the relations between
$\alpha $ and $\gamma $.
For example, when $ \alpha - \gamma \equiv 0 $,
there appears $SU(2)_R$ symmetry associated with
root vectors $\pm (\Theta_3 - \Theta_1)$.
Chiral superfields in the ${\bf 27}$ of $E_6$
are decomposed into the irreducible representations
of $G$.
These situations are summarized as follows;
%
\begin{eqnarray}
    & {\rm (B1)} & \ \ \ \alpha + \gamma \equiv
                       \alpha - \gamma \equiv 0
      \qquad  : \ \ \  G = SU(6) \times SU(2)_R \nonumber \\
     &    & \hspace{30mm} ({\bf 15}, {\bf 1})
                         \ :\ \ Q, g, g^c, L, S  \nonumber \\
     &    & \hspace{30mm} ({\bf 6^*}, {\bf 2})
               \ :\ \ U^c, D^c, H_u, H_d, E^c, N^c \nonumber \\
     & {\rm (B2)} & \ \ \ \alpha + \gamma \equiv 0, \
                       \alpha - \gamma \not\equiv 0
      \ \ : \ \ \ G = SU(6) \times U(1)_R  \nonumber \\
     &    & \hspace{30mm} ({\bf 15}, 0)
                           \ :\ \ Q, g, g^c, L, S  \nonumber \\
     &    & \hspace{30mm} ({\bf 6^*}, 1)
                           \ :\ \ U^c, H_d, N^c \nonumber \\
     &    & \hspace{30mm} ({\bf 6^*}, -1)
                           \ :\ \ D^c, H_u, E^c \nonumber \\
     & {\rm (B3)} & \ \ \ \alpha + \gamma \not\equiv 0, \
                       \alpha - \gamma \equiv 0
      \ \ : \ \ \ G = SU(5) \times SU(2)_R \times U(1)
                                                 \nonumber \\
     &    & \hspace{30mm} ({\bf 10}, {\bf 1}, \frac{2}{3})
                                \ :\ \ Q, g^c, S  \nonumber \\
     &    & \hspace{30mm} ({\bf 5}, {\bf 1}, -\frac{4}{3})
                                \ :\ \ L, g  \nonumber \\
     &    & \hspace{30mm} ({\bf 5^*}, {\bf 2}, -\frac{1}{3})
                       \ :\ \ U^c, D^c, H_u, H_d  \nonumber \\
     &    & \hspace{30mm} ({\bf 1}, {\bf 2}, \frac{5}{3})
                       \ :\ \ E^c, N^c \nonumber \\
     & {\rm (B4)} & \ \ \ \alpha + \gamma \not\equiv 0, \
                       \alpha - \gamma \not\equiv 0
      \ \ : \ \ \ G = SU(5) \times U(1)_R \times U(1)
                                            \nonumber \\
     &    & \hspace{30mm} ({\bf 10}, 0, \frac{2}{3})
                         \ :\ \ Q, g^c, S  \nonumber \\
     &    & \hspace{30mm} ({\bf 5}, 0, -\frac{4}{3})
                         \ :\ \ L, g  \nonumber \\
     &    & \hspace{30mm} ({\bf 5^*}, 1, -\frac{1}{3})
                         \ :\ \ U^c, H_d \nonumber \\
     &    & \hspace{30mm} ({\bf 5^*}, -1, -\frac{1}{3})
                         \ :\ \ D^c, H_u \nonumber \\
     &    & \hspace{30mm} ({\bf 1}, -1, \frac{5}{3})
                         \ :\ \ E^c \nonumber \\
     &    & \hspace{30mm} ({\bf 1}, 1, \frac{5}{3})
                         \ :\ \ N^c \nonumber
\end{eqnarray}
%

When $S$ develops a nonzero VEV,
the gauge group $G$ is spontaneously broken
into a smaller group $G'$.
For each case we have
%
\begin{eqnarray}
     &{\rm (B1)}& \ \ \  G' = SU(4) \times
                         SU(2)_L \times SU(2)_R,  \nonumber \\
     &{\rm (B2)}& \ \ \  G' = SU(4) \times
                         SU(2)_L \times U(1)_R,   \nonumber \\
     &{\rm (B3)}& \ \ \  G' = SU(3)_c \times
                         SU(2)_L \times SU(2)_R \times U(1),
                                                  \nonumber \\
     &{\rm (B4)}& \ \ \  G' = SU(3)_c \times
                         SU(2)_L \times U(1)_R \times U(1). \nonumber
\end{eqnarray}
%
The $SU(4)$ in the cases (B1) and (B2)
is the Pati-Salam $SU(4)$
\cite{PS}.
In the cases (B1) and (B2), $({\bf 4, 2})$, $({\bf 4^*, 2})$
and $({\bf 1, 1})$ gauge superfields under $SU(4) \times SU(2)_L$
absorb  a pair of $Q, L$ and
${\overline Q}, {\overline L}$ and
$(S - {\overline S})/{\sqrt 2}$ via Higgs mechanism.
In the cases (B3) and (B4), $({\bf 3, 2})$, $({\bf 3^*, 2})$
and $({\bf 1, 1})$ gauge superfields under $SU(3)_c \times SU(2)_L$
absorb  a pair of $Q$ and ${\overline Q}$ and
$(S - {\overline S})/{\sqrt 2}$.

In the subsequent symmetry breaking due to
nonzero $\langle N^c \rangle$ the gauge group $G'$
is broken to $G_{st}$.
In the case (B1) a pair of $U^c, E^c$ and
${\overline U^c}, {\overline E^c}$ and
$(N^c - {\overline N^c})/{\sqrt 2}$
are absorbed by gauge superfields.
In the case (B2) a pair of $U^c$ and
${\overline U^c}$ and $(N^c - {\overline N^c})/{\sqrt 2}$
and in the case (B3) a pair of $E^c$ and
${\overline E^c}$ and $(N^c - {\overline N^c})/{\sqrt 2}$
are absorbed.
In the case (B4) only $(N^c - {\overline N^c})/{\sqrt 2}$
is absorbed.


\section{A simple model}

In this section we construct a simple three-generation
model for the case (B4) $G = SU(5)_A \times U(1)^2$.
{}From the observation of this example
we will see that the discrete symmetry of the compactified
manifold controls many parameters of the low-energy
effective theory.
To obtain three-generation models at low energies,
the difference between the generation number and
the anti-generation number should be three at
the Planck scale.
Concretely, here the generation number and
the anti-generation number are taken as 4 and 1,
respectively.
This generation structure is illustrated in Table II.
%
\vskip 0.5cm

\begin{center}
 \unitlength=0.8cm
 \begin{picture}(2.5,2.5)
  \thicklines
  \put(0,0){\framebox(3,1){\bf Table II}}
 \end{picture}
\end{center}

\vskip 0.5cm
%

When the effective theory has Gepner type of discrete
symmetry $Z_{2k+1} \times Z_2$ coming from the symmetry
of the compactified manifold,
nonrenormalizable terms of the superpotential
have peculiar structure.
Especially, if $Z_{2k+1}$-charges of $S_0, \overline S$
and $N^c_0, \overline N^c$ are 1 and $k$, respectively,
the nonrenormalizable terms incorporated only by
these fields are of special form
\cite{majom}
%
\begin{equation}
     W_{NR} \sim \lambda M_C^3 \left[
                  \left(\frac {S_0 {\overline S}}
                                {M_C^2} \right)^{2k}
                + k \left( \frac {N^c_0 {\overline N^c}}
                                  {b^2 M_C^2} \right)^2
                - 2c \left(\frac {S_0 {\overline S}}
                                    {M_C^2} \right)^{k}
                     \left( \frac {N^c_0 {\overline N^c}}
                                       {b^2 M_C^2} \right)
                       \right],
\label{eqn:WNR}
\end{equation}
%
where $M_C$ represents the compactification scale
and $\lambda ,b$ and $c$ are real constants of $O(1)$.
Here we assume that the soft susy-breaking mass parameter
$m_{S_0}^2 + m_{\overline S}^2$,
whose running behavior is controlled by
the renormalization group equation,
becomes negative
in the energy region $O(10^{17})$GeV.
As investigated in Ref.\cite{majom},
carrying out the minimization of the scalar potential
under the conditions $k=3,4,\cdots $ and
$0 <  c < \sqrt{2k}$ $(c \neq \sqrt k)$,
we obtain
%
\begin{eqnarray}
        \langle S_0\rangle  = \langle {\overline S}\rangle
                                       & \sim &  M_C x,  \\
        \langle N^c_0\rangle  = \langle {\overline N^c}\rangle
                                       & \sim &  M_C x^k
\end{eqnarray}
%
with
%
\begin{equation}
     x = \left( \frac {m_{\rm susy}}{M_C} \right)^{1/(4k-2)}.
\end{equation}
%

Through Higgs mechanism $({\bf 3, 2})$ and $({\bf 3^*, 2})$
gauge superfields become massive at the scale
$\langle S_0\rangle  = \langle {\overline S}\rangle $.
It is $Q_0$ and ${\overline Q}$
that are absorbed by $({\bf 3, 2})$ and
$({\bf 3^*, 2})$ gauge superfields.
Since gauge interactions are diagonal with respect to
the generation degree of freedom,
the superfields absorbed here have the same generation indices
as $S_0$ and ${\overline S}$.
Thus at energies below $\langle S_0\rangle  =
\langle {\overline S}\rangle $
only three generations of quark $Q_i(i=1,2,3)$
remain massless.
Through the symmetry breaking $(S_0 - {\overline S})/{\sqrt 2}$
is also absorbed by a gauge superfield associated with
a diagonal generator.
Remaining massless $S$ fields become $S_i(i=1,2,3)$
and $(S_0 + {\overline S})/{\sqrt 2}$.

Leptoquark superfields $g,\ g^c$
$({\overline g},\ {\overline g^c})$ gain their masses of
order $\langle S_0\rangle  = \langle {\overline S}\rangle $
through the Yukawa interactions
$\sum _{i,j=0\sim 3} \lambda ^{(1)}_{i0j}\,g^c_iS_0g_j$
$({\overline \lambda ^{(1)}}\,
{\overline g^c}{\overline S}{\overline g})$.
Here $\lambda ^{(1)}_{i0j}$ can be considered as
a matrix with respect to the indices $i,j$.
If this matrix is rank four,
all $g$ and $g^c$ are massive.
On the other hand, doublet Higgs get their masses
through the Yukawa interactions
$\sum _{i,j=0 \sim 3} \lambda ^{(2)}_{0ij}\,S_0H_{di}H_{uj}$
$({\overline \lambda ^{(2)}}\,{\overline S}
{\overline H_d}{\overline H_u})$.
If the matrix $\lambda ^{(2)}_{0ij}$ is rank three,
a pair of $H_u$ and $H_d$ remains massless and
the other three pairs of them become massive at the scale
$\langle S_0\rangle  = \langle {\overline S}\rangle $.
As seen in Table-II, since $g, g^c, H_u$ and $H_d$
belong to different irreducible representations of
$SU(5)_A$ with each other,
it is likely that the Yukawa couplings
$\lambda ^{(1)}_{i0j}$
and $\lambda ^{(2)}_{0ij}$ have distinct structure
with respect to their ranks.
In the present model triplet-doublet splitting
is attributable to the disparity of ranks of
$\lambda ^{(1)}_{i0j}$ and $\lambda ^{(2)}_{0ij}$.
In Eq.(\ref{eqn:WY}) we have the Yukawa interactions
$\sum \lambda ^{(4)}_{ijk}(L_iH_{uj} + g_iD^c_j)N^c_k$.
In the subsequent symmetry breaking due to
a nonzero $\langle N^c_0 \rangle $
there possibly appear the mixings
between $H_{uj}$ and $L_i^{\dag }$
and between $D^c$ and $g^{\dag }$.
If $\lambda ^{(4)}_{ij0}$'s vanish for all $i$ and $j$,
these mixings are avoidable.
The condition $\lambda ^{(4)}_{ij0} = 0 $
can be explained under
appropriate charge assignments for
the discrete symmetry $Z_{2k+1}\times Z_2$
to $L_i$, $H_{uj}$ and $N^c_0$.

In the present model $U^c, D^c, L$ and $E^c$ also have
four generations and an anti-generation at the Planck scale.
If these superfields have appropriate charges
of the discrete symmetry $Z_{2k+1} \times Z_2$,
we get the nonrenormalizable terms
%
\begin{eqnarray}
  & &  \frac {1}{M_C^{2l_1-1}}(S_0 {\overline S})^{l_1}
                   (U^c_0 {\overline U^c}) +
    \frac {1}{M_C^{2l_2-1}}(S_0 {\overline S})^{l_2}
                   (D^c_0 {\overline D^c}) \nonumber \\
  & & \qquad \qquad  +  \frac {1}{M_C^{2l_3-1}}
                   (S_0 {\overline S})^{l_3}
                   (L_0 {\overline L}) +
    \frac {1}{M_C^{2l_4-1}}(S_0 {\overline S})^{l_4}
                   (E^c_0 {\overline E^c}),
\end{eqnarray}
%
where $l_i < 2k-1 \ (i=1 \sim 4)$.
These terms induce masses for $U^c_0, D^c_0, L_0, E^c_0$
and ${\overline U^c}, {\overline D^c}, {\overline L},
{\overline E^c}$.
In fact, by substituting $S_0, {\overline S}$
by their nonzero VEVs,
we get
%
\begin{eqnarray}
       M_{U^c_0,{\overline U^c}} & \sim & M_C x^{2l_1}, \ \ \ \ \
       M_{D^c_0,{\overline D^c}}  \sim  M_C x^{2l_2}, \nonumber \\
       M_{L_0,{\overline L}}     & \sim & M_C x^{2l_3}, \ \ \ \ \
       M_{E^c_0,{\overline E^c}}  \sim  M_C x^{2l_4}.
\end{eqnarray}
%
Consequently, at energies below $M_{U^c_0,{\overline U^c}}$,
$M_{D^c_0,{\overline D^c}}$, $M_{L_0,{\overline L}}$
and $M_{E^c_0,{\overline E^c}}$
there remain only three generations of
$U^c, D^c, L$ and $E^c$.

At the scale $\langle N^c_0\rangle $,
$(N^c_0 - {\overline N^c})/{\sqrt 2}$ is
absorbed by a gauge superfield.
By assigning appropriate charges to $N^c_i$,
we can obtain large Majorana-masses of $N^c$,
which lead to sufficiently small neutrino masses
by see-saw mechanism.
As shown in Ref.\cite{majom},
the superpotential Eq.(\ref{eqn:WNR}) leads to
Majorana-masses
%
\begin{equation}
         M_M \sim M_C x^{2k} = \sqrt{m_{\rm susy}M_C}\,x
\end{equation}
%
for $N^c_i (i=1,2,3)$ and for
%
\begin{equation}
     N' = \cos \theta \frac {1}{\sqrt 2}(N^c_0 + {\overline N^c})
            + \sin \theta \frac {1}{\sqrt 2}(S_0 + {\overline S})
\end{equation}
%
with
%
\begin{equation}
      \theta \sim x^{k-1}.
\end{equation}
%
Thus at energies below $M_M$ available $G_{st}$-singlet
superfields are limited only to $S_i (i=1,2,3)$ and to
%
\begin{equation}
     S' = - \sin \theta \frac {1}{\sqrt 2}(N^c_0 + {\overline N^c})
            + \cos \theta \frac {1}{\sqrt 2}(S_0 + {\overline S}),
\end{equation}
%
whose masses are $O(m_{\rm susy})$.

As an example, let us consider the case $k = 5$.
In this case the discrete symmetry becomes
$Z_{11} \times Z_2$.
The $Z_{11}$-charges of $S_0, {\overline S}$
and $N^c_0, {\overline N^c}$ are 1 and 5, respectively.
Here we take $l_1 = l_2 = l_3 = l_4 = 3 $
and take numerical values of $M_C$ and $m_{\rm susy}$
as
%
\begin{equation}
      M_C \cong \frac {M_{\rm Pl}}{\sqrt {8\pi }}
                   \cong 10^{18.4}{\rm GeV}, \ \ \ \
                       m_{\rm susy} = 10^3{\rm GeV}.
\end{equation}
%
In this case we get $x = 10^{-0.86}$
and mass hierarchies become
%
\begin{eqnarray}
   & &   \langle S_0\rangle  = \langle {\overline S}\rangle
                  \cong  10^{17.5}{\rm GeV}, \\
   & &   \langle N^c_0\rangle  = \langle {\overline N^c}\rangle
                   \cong  10^{14.1}{\rm GeV}, \\
   & &   M_{U^c_0,{\overline U^c}},
           \ M_{D^c_0,{\overline D^c}},
             \ M_{L_0,{\overline L}},
               \ M_{E^c_0,{\overline E^c}}
                          \cong  10^{13.3}{\rm GeV}, \\
   & &   M_M  \cong  10^{9.8}{\rm GeV}.
\end{eqnarray}
%
Large Majorana-masses $M_M$ obtained here
solve the solar neutrino problem
\cite{majom}.
At energies below $M_M$
this model is in accord with the minimal
supersymmetric standard model except for
the existence of singlet fields
$S_i (i=1,2,3)$ and $S'$.

Now it is interesting to study the unification
of gauge coupling constants.
In the aligned $SU(5) \times U(1)^2$ model,
$SU(3)_c$ and $SU(2)_L$ gauge couplings should
be unified at the scale $\langle S \rangle $
but not at the Planck scale.
On the other hand, due to possible existence of
gauge kinetic mixing terms unification of
abelian gauge couplings is not straightforward
\cite{mixing}.
Here we confine ourselves to non-abelian
gauge couplings.
The one-loop renormalization group
equation for gauge couplings reads
%
\begin{equation}
     \frac {d\alpha _i}{dt} = \frac {1}{2 \pi}
             b_i \alpha_i^2 \ \ \ \ \qquad (i = 3, 2)
\end{equation}
%
with $t = \ln (\mu /\mu _0)$.
In the model explored above the coefficients of
$\beta $-functions for $SU(3)_c$ and $SU(2)_L$
gauge couplings are given by
%
\begin{equation}
      b_3 = -1, \ \ \ \ \ b_2 = 2
\end{equation}
%
in the energy region from $\langle S \rangle $
to $M_{U^c_0,{\overline U^c}}$ and
%
\begin{equation}
      b_3 = -3, \ \ \ \ \ b_2 = 1
\end{equation}
%
in the energy region from $M_{U^c_0,{\overline U^c}}$
to $m_{\rm susy}$.
Therefore, for each region the difference $b_2 - b_3$
becomes 3 and 4, respectively.
To the contrary, in the minimal supersymmetric
standard model the difference is equal to 4
over the range from $M_{\rm GUT}$ to $m_{\rm susy}$.
The present model leads to the relation
%
\begin{equation}
     \alpha _2(M_Z)^{-1} - \alpha _3(M_Z)^{-1}
                   =  \frac {1}{4\pi }\left[
                   8\,\ln \left( \frac {M_C}{M_Z}\right)
                   - \frac {2l+3}{2k-1} \,
                   \ln \left( \frac {M_C}{m_{\rm susy}}\right)
                   \right]
\end{equation}
%
in the one-loop renormalization group calculation,
where $l = l_1 = l_2 = l_3 = l_4 $.
We use the unification condition
$\alpha _3 = \alpha _2$ at the scale $\langle S \rangle$
and Eqs.(19) and (21).
As far as the difference $\alpha _2(\mu )^{-1} -
\alpha _3(\mu )^{-1}$ is concerned,
the two-loop effect gives
only a small correction to the one-loop effect.
After numerical calculations we find that
when $l = k-2 $,
the unification of $SU(3)_c$ and $SU(2)_L$
gauge couplings at the scale $\langle S \rangle $
is consistent with experimental data.
Detailed renormalization group analysis of
gauge couplings including abelian ones
will be presented elsewhere.


\section{Summary and Discussion}

In Calabi-Yau string compactification,
there possibly exist three kinds of
$SU(5) \times U(1)^2$ gauge symmetry which
contain $SU(3)_c \times SU(2)_L \times U(1)_Y$.
Among them realistic models can be constructed
only in the case of the aligned $SU(5) \times U(1)^2$
gauge symmetry,
in which the $SU(5)$ differs from the standard
$SU(5)$ and also from the flipped $SU(5)$.
In this model the gauge group
$G = SU(5)_A \times U(1)^2$ at the Planck scale
is spontaneously broken
into $G_{st}$ by two stages
when $G_{st}$-neutral fields in
the ${\bf 27}$ of $E_6$ develop
nonzero VEVs.
At the first stage, when the field $S$ in ${\bf 10}$ of
$SU(5)_A$ evolves its VEV of $O(\gsim 10^{16})$GeV,
$G$ is broken into
%
\begin{equation}
    G' = SU(3)_c \times SU(2)_L \times U(1)^2.
\end{equation}
%
Subsequent symmetry breaking from $G'$ to $G_{st}$
is attributed to a nonzero VEV of $N^c$.
Although the unification scale of all
fundamental interactions is the Planck scale,
$SU(3)_c$ and $SU(2)_L$ gauge couplings
come together at the scale $\langle S \rangle
= O(10^{17.5})$GeV.
Therefore, the string threshold effect takes part in
the unification of $SU(5)_A$ and $U(1)^2$ gauge
couplings but does not in the unification of
$SU(3)_c$ and $SU(2)_L$ gauge couplings.

In this paper we constructed a simple
three-generation model with
the aligned $SU(5) \times U(1)^2$.
Under appropriate charge assignments
of Gepner type of discrete symmetry
mass spectra of the model comes down to as follows.
Through Higgs mechanism at the scale $\langle S_0\rangle $,
chiral superfields $Q_0, {\overline Q}$ and
$(S_0 - {\overline S})/{\sqrt 2}$
are absorbed by gauge superfields.
At the same scale all of $g, g^c$ and ${\overline g},
{\overline g^c}$ become massive.
All but one set of $H_u$ and $H_d$ also gain their masses.
At the next stage of symmetry breaking
due to $\langle N^c_0\rangle $,
$(N^c_0 - {\overline N^c})/{\sqrt 2}$ is absorbed.
Chiral superfields $U^c_0, D^c_0, L_0, E^c_0$
and ${\overline U^c}, {\overline D^c}, {\overline L},
{\overline E^c}$ get masses of order $M_C\,x^{2l_i}$
through the nonrenormalizable interactions.
Chiral superfields $N^c_i (i=1,2,3)$ and
$N'(\sim N^c_0 + {\overline N^c})$
get large Majorana-masses $M_M \sim M_C\,x^{2k}
= \sqrt {m_{\rm susy}M_C} \,x$
also via nonrenormalizable interactions.
Thus, the triplet-doublet splitting problem and
solar neutrino problem can be solved with
the aid of the discrete symmetry.
Consequently, at energies below $M_M$,
$M_{U^c_0,{\overline U^c}}$,
$M_{D^c_0,{\overline D^c}}$, $M_{L_0,{\overline L}}$
and $M_{E^c_0,{\overline E^c}}$
available superfields
are reduced to three generations of
$Q_i, U^c_i, D^c_i, L_i, E^c_i (i=1,2,3)$,
a pair of Higgs superfield $H_u, H_d$
and singlet superfields $S_i (i=1,2,3),
S'(\sim S_0 + {\overline S})$.
The model obtained here is in accord with the minimal
supersymmetric standard model except for
the existence of singlet fields
$S_i (i=1,2,3)$ and $S'$ with masses of $O(m_{\rm susy})$.

In the present model we can find a realistic
solution also for the $\mu $-problem.
Since there is a nonrenormalizable term
%
\begin{equation}
  \frac {1}{M_C^{2n}}(S_0 {\overline S})^n S_0 H_u H_d
\end{equation}
%
for a pair of light Higgs fields $H_u$ and $H_d$,
we obtain the induced $\mu $-term with
%
\begin{equation}
    \mu \sim M_C\,x^{2n+1}.
\end{equation}
%
If the sum of $Z_{2k+1}$-charges of $H_u$ and $H_d$
is 1,
then we get $n = 2k-1$ and
%
\begin{equation}
    \mu \sim m_{\rm susy}\, x.
\end{equation}
%
By taking $m_{\rm susy} \sim $ 1TeV and $k = 5$,
one finds
%
\begin{equation}
     \mu \sim 100{\rm GeV}.
\end{equation}
%
This is a plausible solution for the $\mu $-problem.
Moreover, there is a possibility that
the present model gives a plausible interpretation
of quark/lepton mass hierarchy.
The problem will be studied in detail elsewhere
\cite{ours}.
The discrete symmetry of the compactified manifold
as well as the supersymmetry breaking and
the gauge hierarchy plays an important role
in connecting the superstring theory with
the standard model and in determining the parameters
of the standard model.



\newpage

{\bf Table Captions}

\vspace {1cm}

{\bf Table I}\ \ Irreducible decompositions of the {\bf 27}
matter superfields under three kinds of $SU(5) \times U(1)^2$.
$U(1)^2$-axes in three cases correspond to
$\Theta _2 \mp \Theta _3$, $\Theta _3 \mp \Theta _1$
and $\Theta _1 \mp \Theta _2$, respectively.
The numbers in parentheses are the dimensions
of the $SU(5)$ representations and
the quantum numbers of $U(1)^2$.

\vspace {1cm}

{\bf Table II}\ \ The generation and anti-generation
structure of matter superfields in a simple
three-generation model with the aligned $SU(5) \times U(1)^2$.
$U(1)^2$-axes correspond to $\Theta _3 - \Theta _1$
and $\Theta _3 + \Theta _1$.


\newpage

\begin{center}
{\bf Table I }\\
\vspace {5mm}

\begin{tabular}{|l|l|l|l|}  \hline
$SU(5) \times U(1)^2$ & (A) \ \ $SU(5)_S$ &
                         (B) \ \ $SU(5)_A$ & (C) \ \ $SU(5)_F$   \\
\hline  \hline
$({\bf 10},\ \ 0,\ \ \frac{2}{3})$    & $Q, \ U^c, E^c$  & $Q, \ g^c, S$
                                            & $Q, \ D^c, N^c$  \\
$({\bf 5^*},\ \ 1,-\frac{1}{3})$      & $D^c, L$         & $U^c, H_d$
                                            & $g^c, H_u$     \\
$({\bf 5^*},-1,-\frac{1}{3})$         & $g^c, H_d$       & $D^c, H_u$
                                            & $U^c, L$       \\
$(\ {\bf 5},\ \ 0,-\frac{4}{3})$      & $g, \ H_u$       & $L, \ g$
                                            & $g, \ H_d$       \\
$(\ {\bf 1},-1,\ \ \frac{5}{3})$      & $N^c$            & $E^c$
                                            & $S$            \\
$(\ {\bf 1},\ \ 1,\ \ \frac{5}{3})$   & $S$              & $N^c$
                                            & $E^c$          \\
\hline
\end{tabular}

\end{center}

\vspace {3cm}

\begin{center}
{\bf Table II }\\
\vspace {5mm}

\begin{tabular}{|l|l r|l|}  \hline
$SU(5)_A \times U(1)^2$  &  \ \ \ \ \ \ generation  &
                                      &  anti-generation  \\
\hline  \hline
$({\bf 10},\ \ 0,\ \ \frac{2}{3})$     &  $(Q, \ g^c, S)_i$ &  $(i=0,1,2,3)$
          &  $({\overline Q}, \ {\overline g^c}, {\overline S})$   \\
$({\bf 5^*},\ \ 1,-\frac{1}{3})$  &  $(U^c, H_d)_i$  &  $(i=0,1,2,3)$
              &   $({\overline U^c}, {\overline H_d})$    \\
$({\bf 5^*},-1,-\frac{1}{3})$   &  $(D^c, H_u)_i$  &  $(i=0,1,2,3)$
              &   $({\overline D^c}, {\overline H_u})$    \\
$(\ {\bf 5},\ \ 0,-\frac{4}{3})$     &  $(L, \ g)_i$      &  $(i=0,1,2,3)$
              &   $({\overline L}, \ {\overline g})$        \\
$(\ {\bf 1},-1,\ \ \frac{5}{3})$      &  $(E^c)_i$       &  $(i=0,1,2,3)$
              &   $({\overline E^c})$    \\
$(\ {\bf 1},\ \ 1,\ \ \frac{5}{3})$     &  $(N^c)_i$       &  $(i=0,1,2,3)$
              &   $({\overline N^c})$                     \\
\hline
\end{tabular}

\end{center}


\end{document}